\documentclass[aps,prl,twocolumn,showkey,square,numbers,amssymb,amsmath,nobibnotes,footinbib]{revtex4}
\usepackage{esint}
\usepackage{bm}
\usepackage{amsmath}
\usepackage{amsfonts}
\usepackage{amssymb}  
\usepackage{verbatim}
\usepackage{times}
\usepackage{graphicx}
\usepackage{dsfont}
\usepackage{titletoc} 
\usepackage{microtype}

\DeclareBoldMathCommand\boldlangle{\left\langle}
\DeclareBoldMathCommand\boldrangle{\right\rangle}

\newcommand{\barr}{\begin{eqnarray}}
\newcommand{\earr}{\end{eqnarray}}
\newcommand{\beq}{\begin{equation}}
\newcommand{\eeq}{\end{equation}}
\newcommand{\be}{\begin{equation}}
\newcommand{\ee}{\end{equation}}

\newcommand{\de}{\mathrm{d}}

\newcommand{\avg}[1]{\left< #1 \right>} 
 
\let\baraccent=\= 
\renewcommand{\=}[1]{\stackrel{#1}{=}} 

\newcommand{\numberset}{\mathbb}

\newcommand{\R}{\numberset{R}}

\usepackage{graphicx,color}

\begin{document}


\title{Universal covariance formula for linear statistics on random matrices}

\author{Fabio Deelan Cunden$^{1,2}$ and Pierpaolo Vivo$^{3}$}
\affiliation{$1$. Dipartimento di Matematica, Universit\`a di Bari, I-70125 Bari, Italy\\
$2$. Istituto Nazionale di Fisica Nucleare (INFN), Sezione di Bari, I-70126 Bari, Italy\\
$3$. Laboratoire de Physique Th\'{e}orique et Mod\`{e}les
Statistiques (UMR 8626 du CNRS), Universit\'{e} Paris-Sud,
B\^{a}timent 100, 91405 Orsay Cedex, France}
\date{\today}

\begin{abstract} 
We derive an analytical formula for the covariance $\mathrm{Cov}(A,B)$ of two smooth linear statistics $A=\sum_i a(\lambda_i)$ and $B=\sum_i b(\lambda_i)$ to leading order for $N\to\infty$, where $\{\lambda_i\}$ are the $N$ real eigenvalues of a general one-cut random-matrix model with Dyson index $\beta$. The formula, carrying the universal $1/\beta$ prefactor, depends on the random-matrix ensemble only through the edge points $[\lambda_-,\lambda_+]$ of the limiting spectral density. For $A=B$, we recover in some special cases the classical variance formulas by Beenakker and Dyson-Mehta, clarifying the respective ranges of applicability. Some choices of $a(x)$ and $b(x)$ lead to a striking \emph{decorrelation} of the corresponding linear statistics. We provide two applications - the joint statistics of conductance and shot noise in ideal chaotic cavities, and some new fluctuation relations for traces of powers of random matrices. 
\end{abstract}


\maketitle

\textit{Introduction - } The discovery of the phenomenon of \emph{universal conductance fluctuations} (\textsf{UCF}) in disordered metallic samples, pioneered by Altshuler \cite{alt} and Lee and Stone \cite{leestone} has had a profound impact on our current understanding of the mechanisms of quantum transport at low temperatures and voltage. There are two aspects of this universality, $\mathrm{i}.)$ the variance of the conductance is of order $(e^2/h)^2$, \emph{independent of} sample size or disorder strength, and $\mathrm{ii}.)$
this variance decreases by precisely a factor of two if time-reversal symmetry is broken by a magnetic field. Both features, observed in several experiments and numerical simulations (see \cite{benreview} for a review), naturally emerge from a random-matrix theoretical formulation of the electronic transport problem \cite{jal1,mello}. The phenomenon of \textsf{UCF} is just, however, one of the very many incarnations of a more general and intriguing property of sums of strongly correlated random variables.

Consider first, for instance, a set of $N$ i.i.d. $\mathcal{O}(1)$ random variables $\{X_i\}$. The random variable $A=\sum_{i} a(X_i)$, for any function $a(x)$ (hereafter all summations run from $1$ to $N$), is called a \emph{linear statistics} of the sample $\{X_i\}$. 
For large $N$, both the average $\avg{A}$ and the variance $\mathrm{Var}(A)$ typically grow linearly with $N$. But what happens if the $N$ variables are instead strongly correlated? A prominent example is given by the $N$ real eigenvalues $\{\lambda_i\}$ of a random matrix. In this case a completely different behavior emerges: if $a(x)$ is sufficiently smooth \footnote{It is sufficient to have $a(x)$ twice-differentiable. If $a(x)$ is non-smooth, $\mathrm{Var} (A)$ typically grows logarithmically with $N$.}, while the average is still of order $\mathcal{O}(N)$, the variance attains a \emph{finite} value for $N\to\infty$. Moreover, quite generally $\mathrm{Var}(A)\propto 1/\beta$, where $\beta$ (the Dyson index) is related to the symmetries of the ensemble, and on the scale $\mathcal{O}(1)$ of typical fluctuations around the average, the distribution of $A$ is Gaussian \cite{politzer,chenmanning,forr1,sosh1,lytova,tracy,pastur,johansson,costin,forrleb}. Recalling that the conductance in chaotic cavities can be indeed written as a linear statistics of a random matrix (see below), the phenomenon of \textsf{UCF} is readily understood. The issue of fluctuations of generic linear statistics has however a longer history in the physics and mathematics literature \cite{politzer,chenmanning,forr1,sosh1,lytova,pastur,johansson,tracy,costin,forrleb}, due to its relevance for a variety of applications beyond \textsf{UCF}, ranging from quantum transport in metallic conductors \cite{beenakkerreview} and entanglement of trapped fermion chains \cite{vicari} to the statistics of extrema of disordered landscapes \cite{scard} - to mention just a few.

For a smooth $a(x)$, there exist two celebrated formulas in the physics literature by Dyson-Mehta (\textsf{DM}) \cite{dysonmehta}  and Beenakker (\textsf{B}) \cite{beenakkerPRL,beenakkerPRB} for $\mathrm{Var}(A)$, the latter precisely derived in the context of the quantum transport problem introduced earlier (see also \cite{beenakker2} for a generalized \textsf{B} formula). They are deemed \emph{universal} - not dependent on the microscopic detail of the random matrix ensemble under consideration - and correctly predict a $\mathcal{O}(1)$ value for $N\to\infty$ and a universal $\beta^{-1}$ prefactor. 

What happens now if \emph{two} linear statistics $A(\bm\lambda)=\sum_{i} a(\lambda_i)$ and $B(\bm\lambda)=\sum_{i} b(\lambda_i)$ are simultaneously considered? Motivated by applications to the quantum transport problem \cite{cunden} and multivariate data analysis \cite{cunden2}, we set for ourselves the task to find a universal formula for the \emph{covariance} $\mathrm{Cov}(A,B)$ that would reduce to \textsf{DM} \emph{or} \textsf{B} for $A\equiv B$. But before proceeding, it felt natural to first check under which precise conditions should we expect to recover one formula \emph{or} the other.

Much to our surprise, we have failed to find a sufficiently transparent (at least to our eyes) account that encompasses all possible cases in an accessible and systematic way. The goal of this Letter is thus to produce a so-far unavailable universal formula for $\mathrm{Cov}(A,B)$ of large dimensional random matrices. As a byproduct of our result, we generalize \textsf{DM} and \textsf{B} formulas for $A=B$. We introduce a ``conformal map" method which encloses all possible cases (old and new) into a neat and unified framework. We further employ our formula to probe a quite interesting phenomenon of \emph{decorrelation}, namely for some choices of $a(x)$ and $b(x)$ we get $\mathrm{Cov}(A,B)=\mathcal{O}(N^{-1})$. Examples are given for $\mathrm{i}.)$ conductance and shot noise in ideal chaotic cavities supporting a large number of electronic channels, and $\mathrm{ii}.)$ fluctuation relations for traces of powers of random matrices.

\begin{figure}[t]
\centering
\includegraphics[width=1\columnwidth]{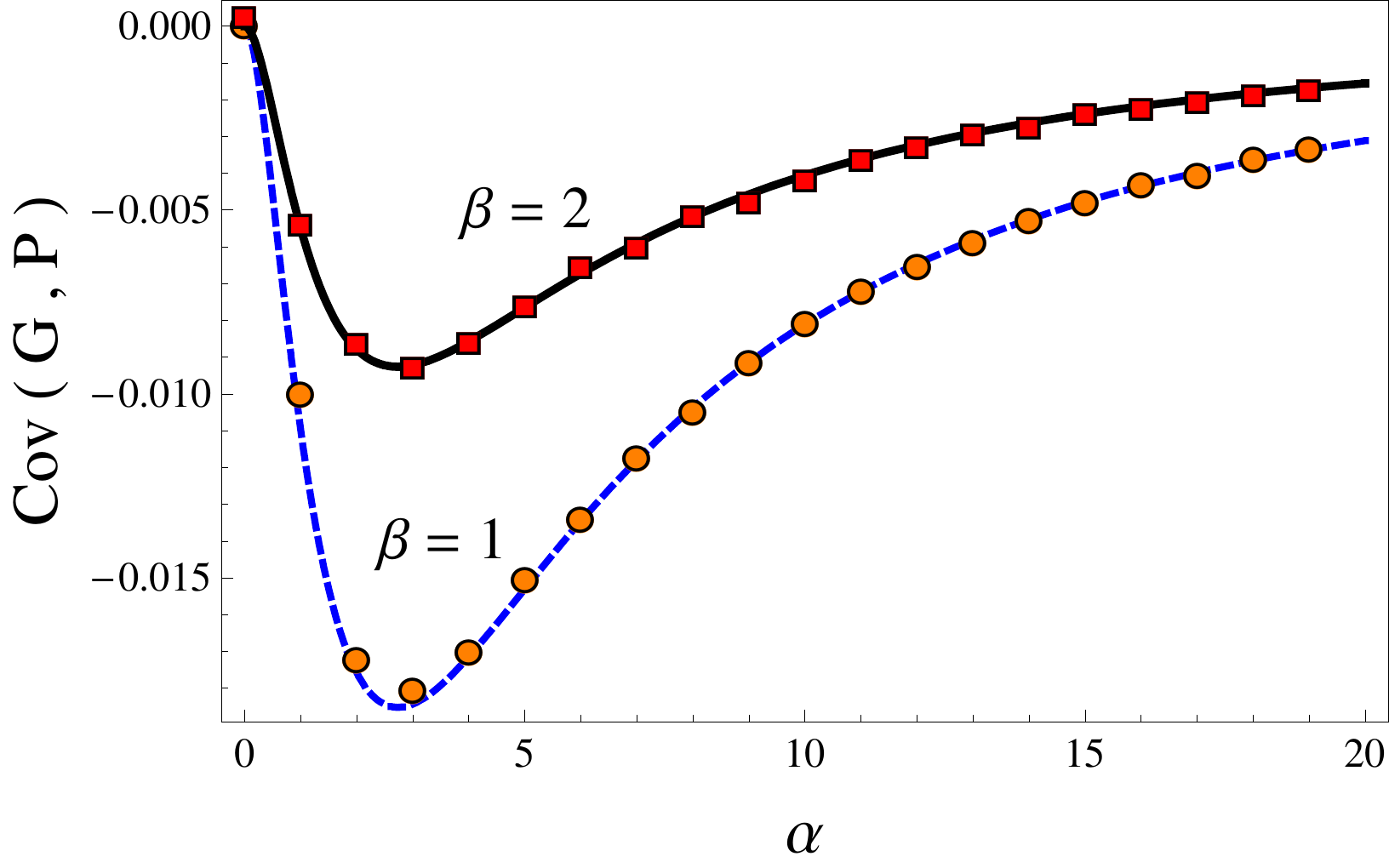}
\caption{(color online) Covariance of the dimensionless conductance $G$ and shot-noise $P$ \eqref{covGP} as a function of $\alpha$. The dashed blue ($\beta=1$) and the solid black ($\beta=2$) lines are the analytical result \eqref{covGP} found using \eqref{cov_fourier}. The points are obtained from a numerical diagonalization of $n=10^{4}$ random Jacobi matrices of size $N=30$. For large $N$ the covariance is zero for symmetric cavities ($\alpha=0$) and vanishes in the limit of high asymmetry ($\alpha\to\infty$). The maximal anticorrelation $\mathrm{Cov}(G,P)=-1/54\beta$ is realized at $\alpha^{\star}=1+\sqrt{3}\approx 2.73205...$, independent of $\beta$. The analytical curve is well reproduced even for the moderate size $N=30$ used in numerical simulations.}
\label{fig:cov}
\end{figure}

\textit{Setting and results - } We consider an ensemble of $N\times N$ random matrices $\mathcal{H}$, whose joint probability density (jpd) of the $N$ eigenvalues $\lambda_i\in\Lambda$ (a generic interval of the real line) can be cast in the Gibbs-Boltzmann form
\be
\mathcal{P}_\beta(\bm{\lambda})=  \frac{1}{\mathcal{Z}}\mathrm{e}^{-\beta \left[-\sum_{i< j}{\ln{|\lambda_i-\lambda_j|}}+N\sum_{i}{V(\lambda_i)}\right]} \equiv \frac{\mathrm{e}^{-\beta E(\bm\lambda)}}{\mathcal{Z}}  \ . \label{jpdf}
\ee

Here, the normalization constant $\mathcal{Z}=\int_{\Lambda^N}{\de\bm{\lambda}\,\mathrm{e}^{-\beta E(\bm{\lambda}) }}$ is the partition function of a Coulomb gas, namely a 1D system of $N$ particles in equilibrium at inverse temperature $\beta>0$ (the Dyson index), whose energy $E(\bm\lambda)$ contains a logarithmic repulsive interaction and a confining single-particle potential $V(x)$. 
We first define the spectral density $\rho_N(\lambda)=N^{-1}\sum_{i}\delta(\lambda-\lambda_i)$ (a random measure on the real line), and its average for finite ($\avg{\rho_N(\lambda)}$) and large $N$ ($\rho(\lambda)=\lim_{N\to\infty}\avg{\rho_N(\lambda)}$), where henceforth $\avg{\cdot}$ stands for averaging with respect to \eqref{jpdf}. The potential $V(x)$ is assumed to be such that $\rho(\lambda)$ is supported on a \emph{single} interval $\sigma$ of the real line (possibly unbounded). 

The form of the jpd \eqref{jpdf} includes classical invariant ensembles \cite{Mehta} such as Wigner-Gauss
$\mathcal{G}$, Wishart-Laguerre $\mathcal{W}$, Jacobi $\mathcal{J}$ and Cauchy $\mathcal{C}$. The first two ensembles are defined as $\mathcal{G}= \frac{\mathcal{Y}+\mathcal{Y}^{\dagger}}{\sqrt{2N}}$ and $\mathcal{W}= \frac{\mathcal{Y}^{\dagger}\mathcal{Y}}{N}$, where $\mathcal{Y}$ is a $M\times N$ random matrix with standard Gaussian independent entries \footnote{Hereafter the elements of $\mathcal{Y}$ are real, complex or real quaternion independent random variables with Gaussian densities $\frac{1}{\sqrt{2\pi}}\mathrm{e}^{-x^2_{ij}/2}$, $\frac{1}{\pi}\mathrm{e}^{-|z_{ij}|^2}$ and $\left(\frac{2}{\pi}\mathrm{e}^{-2|z_{ij}|^2},\frac{2}{\pi}\mathrm{e}^{-2|w_{ij}|^2}\right)$ for $\beta=1,2$ and $4$ respectively (recall that a real quaternion number is specified by two complex numbers $(z,w)$). The symbol $^{\dagger}$ stands
for transpose, hermitian conjugate and symplectic conjugate respectively - yielding a corresponding Dyson index $\beta=1,2,4$.}, with $M = N$ and $M = (1+\alpha)N$
($\alpha\geq0$) for $\mathcal{G}$ and $\mathcal{W}$ respectively. A Jacobi matrix $\mathcal{J}=(\mathcal{W}_1+\mathcal{W}_2)^{-1}\mathcal{W}_1$ is defined in terms of two independent Wishart matrices of parameters $\alpha_{1,2}$.
Finally, the Cauchy ensemble $\mathcal{C}=\mathrm{i}\left(1-\mathcal{U}\right)\left(1+\mathcal{U}\right)^{-1}$ is obtained by a Cayley transform on Haar-distributed unitary matrices $\mathcal{U}$. In Table \ref{table1} the corresponding potentials are listed.  We stress, however, that the general setting in \eqref{jpdf} applies equally well e.g. to \emph{non}-invariant ensembles such as the Dumitriu-Edelman \cite{dumitriu} tridiagonal $\beta$-ensembles, for non-quantized $\beta>0$.
\begin{table}
 \begin{tabular}{||c | c c c||} 
 \hline
 \qquad & $V(x)$ & $\Lambda$ & $\sigma$ \\ [.1ex] 
 \hline\hline
$\mathcal{G}$ & $\frac{x^2}{4}$ & $(-\infty,\infty)$ & $\pm2$ \\ 
 \hline
$\mathcal{W}$ & $\frac{x}{2}-\ln x^{\frac{\alpha}{2}}$ & $[0,\infty)$ & $\left(1\pm\sqrt{1+\alpha}\right)^2$ \\
 \hline
 $\mathcal{J}$ & $-\ln x^{\frac{\alpha_1}{2}}(1-x)^{\frac{\alpha_2}{2}}$ & $[0,1]$ &
 $\left(\frac{\sqrt{1+\alpha_2}\pm\sqrt{(\alpha_1+1)(\alpha_1+\alpha_2+1)}}{\alpha_1+\alpha_2+2}\right)^2$ \\
 \hline
 $\mathcal{C}$ &  $\ln{\sqrt{1+x^2}}$ & $(-\infty,\infty)$ & $(-\infty,\infty)$ \\
 \hline
\end{tabular}
\caption{Summary of various classical ensembles of type \eqref{jpdf}. For $\mathcal{G},\mathcal{W},\mathcal{J}$ we provide the edges of the limiting support $\sigma=[\lambda_{-},\lambda_{+}]$.}\label{table1}
\end{table}

Consider now two linear statistics $A(\bm{\lambda})=\sum_{i} a(\lambda_i)$ and $B(\bm{\lambda})=\sum_{i} b(\lambda_i)$. Their covariance is given by the $N$-fold integral
\be
\mathrm{Cov}(A,B)=\int_{\Lambda^N}\hspace{-1mm}\de\bm{\lambda}\mathcal{P}_\beta(\bm{\lambda})\left(A(\bm{\lambda})-\avg{A}\right)\left(B(\bm{\lambda})-\avg{B}\right)\ .\label{defcov11}
\ee
For smooth $a(x)$ and $b(x)$ we show that this covariance \eqref{defcov11} has the universal form
\be
\mathrm{Cov}(A,B)=\frac{1}{\beta\pi^2} \int_{0}^{\infty} \de k\ \varphi(k)\, \mathrm{Re}\left[\tilde{a}(k)\tilde{b}^{\star}(k)\right]\ , \label{cov_fourier}
\ee
with an error term of order $\mathcal{O}(N^{-1})$, which will always be neglected henceforth. Here $\mathrm{Re}$ stands for the real part and $^\star$ for complex conjugation. Assume that at least one of the end points of $\sigma$ is finite, as in many practical cases. 
Then $\varphi(k)=k\tanh(\pi k)$ is a \emph{universal} kernel and we have introduced a deformed Fourier transform $\tilde{f}(k)=\int_{-\infty}^{+\infty}\de x\, \mathrm{e}^{\mathrm{i}kx} f(T(\mathrm{e}^x))$, where $T(\cdot)$ is a conformal map defined by the edges of the support of $\rho(\lambda)$
\be
T(x)=\begin{cases}
\frac{x\lambda_{-}+\lambda_{+}}{x+1}&\text{for }\sigma=[\lambda_-,\lambda_+]\\
\lambda_- +1/x &\text{for }\sigma=[\lambda_-,\infty)\\
\lambda_+-x &\text{for }\sigma=(-\infty,\lambda_+]\ .\label{T}
\end{cases}
\ee
The role of $T(\cdot)$ is to map the positive  half-line $[0,+\infty)$ to the support $\sigma$ of $\rho(\lambda)$. Since no such conformal mapping exists if $\sigma=(-\infty,+\infty)$, this (unfrequent) case (e.g. the Cauchy ensemble $\mathcal{C}$) must be treated differently. In this case $\varphi(k)=k$ and $\tilde{f}(k)=\int_{-\infty}^{+\infty}\de x\, \mathrm{e}^{\mathrm{i}kx} f(x)$ is the standard Fourier transform. Eq. \eqref{cov_fourier} may be used whenever the integral converges. 

Let us now offer a few remarks. First, formula \eqref{cov_fourier} is evidently symmetric upon the exchange $A\leftrightarrow B$, as $\mathrm{Cov}(A,B)=\mathrm{Cov}(B,A)$. Second, the only dependence on the Dyson index $\beta$ is through the prefactor $\beta^{-1}$ as already anticipated. Third, the details of the confining potential $V(x)$ only appear in the formula \eqref{cov_fourier} through the edges $\lambda_\pm$ of the limiting spectral density $\rho(\lambda)$, \emph{and not} through the range of variability of the eigenvalues $\Lambda$ \footnote{For example, for the Gaussian ensemble the support of the jpd $\mathcal{P}_{\beta}(\bm{\lambda})$ is $\Lambda^N$ with $\Lambda=(-\infty,+\infty)$, \emph{but} the average density $\rho(\lambda)$ is the Wigner's law, supported on $\sigma=[-2,2]$. Therefore one should use the kernel $\phi(k)=k\tanh{\left(\pi k\right)}$, with the conformal map \eqref{T} defined by $\lambda_\pm=\pm2$.}. This is a consequence of universality of the (smoothed) two-point kernel \cite{ambjorn,brezinzee,brezinzee2,kanz}. Fourth, if $\sigma=[\lambda_-,\lambda_+]$, the covariance admits the following alternative expression in real space
\be
\mathrm{Cov}(A,B)=\frac{1}{\beta\pi^2}P\hspace{-1mm}\iint_{\lambda_-}^{\lambda_+}\!\!\!\!\!\ \de\lambda\de\lambda^\prime \phi(\lambda,\lambda^\prime)\frac{a(\lambda^\prime)}{\lambda^\prime-\lambda}\frac{\de b(\lambda)}{\de\lambda}\ ,\label{alternativecov} 
\ee 
where $\phi(\lambda,\lambda^\prime)=\sqrt{\frac{(\lambda_+ -\lambda)(\lambda-\lambda_-)}{(\lambda_+ -\lambda^\prime)(\lambda^\prime -\lambda_-)}}$ and $P$ stands for Cauchy's principal value. Formula \eqref{alternativecov}, which may be more convenient than \eqref{cov_fourier} in certain cases, reduces for $a(x)=b(x)$ to the generalized \textsf{B} formula for the variance (as given in \cite{beenakker2}, Eq. (17)). On the other hand, \eqref{cov_fourier} recovers for $a(x)=b(x)$ the \textsf{DM} formula \cite{dysonmehta} (see Eq. (1.1) in \cite{beenakkerPRB}) if $\sigma=(-\infty,+\infty)$, and the \textsf{B} formula \cite{beenakkerPRL} (see \eqref{formulaB} below) if $\sigma=[0,1]$. Eq. \eqref{cov_fourier} and \eqref{T} constitute then a neat and unified summary of all possible occurrences, including the case of semi-infinite supports (relevant for some cases \cite{burda,wieczorek}). Fifth, the representation \eqref{cov_fourier} in Fourier space makes apparent that the covariance \emph{vanishes} to leading order e.g. if $\tilde{a}(k)$ is purely imaginary and $\tilde{b}(k)$ is real. Consider for instance a case with an even potential $V(x)=V(-x)$, like the Wigner-Gauss $\mathcal{G}$. Then, if the linear statistics $A$ is defined by an even function $a(-y)=a(y)$, its deformed Fourier transform $\tilde{a}(k)$ is real, while if $a(-y)=-a(y)$, then $\tilde{a}(k)$ is purely imaginary \footnote{In this case $\lambda_{-}=-\lambda_{+}$ and $T(x)=\lambda_{+}\left(1-x\right)/\left(1+x\right)$. It is immediate to verify that $T(\mathrm{e}^{-x}) = - T(\mathrm{e}^x)$.}. This simple observation immediately predicts that the moments $\mathrm{Tr} {\,\mathcal{G}^n}$ of a Gaussian matrix (or any random matrix with an even potential) are asymptotically pairwise uncorrelated $\mathrm{Cov}(\mathrm{Tr} {\,\mathcal{G}^n},\mathrm{Tr} {\,\mathcal{G}^m})=\mathcal{O}(N^{-1})$ if $n$ is even and $m$ odd. We provide now two examples of applications of the covariance formula, before turning to its derivation.

\textit{Examples - } As a first example, we focus on the random matrix theory of quantum transport as discussed in the introduction. At low temperature and voltage, the electronic transport in mesoscopic cavities whose classical dynamics is chaotic can be modeled by a scattering matrix $\mathcal{S}$ of the system uniformly distributed in the unitary group \cite{jal1,mello} (for a review see \cite{beenakkerreview} and references therein). The $M\times M$ matrix $\mathcal{S}$
is just unitary if time-reversal symmetry is broken $(\beta=2)$, or unitary and symmetric in case of preserved time-reversal symmetry $(\beta=1)$, where unitarity is required by charge conservation. The size $M=N_1+N_2$ is determined by the number $N_{1,2}$ of open channels in the two leads attached to the cavity, and we denote $N=\min(N_1,N_2)$. Many experimentally accessible quantities can be expressed as linear statistics of the form $A=\sum_{i} a(\lambda_i)$, where $\lambda_i\in [0,1]$ are so-called transmission eigenvalues. They are the eigenvalues of the (random) hermitian matrix $\mathcal{T}=t t^\dagger$, where $t$ is a $\min(N_{1,2})\times\max(N_{1,2})$ submatrix of $\mathcal{S}$ and $a(x)$ is an appropriate function. 
For instance, as already disclosed in the introduction, the dimensionless conductance $G$ and shot noise $P$ \footnote{The conductance $G$ is measured in units of the conductance quantum $G_0 = 2e^2/\hbar$, where $e$ is the electronic charge and $\hbar$ is Planck's constant, while the shot noise $P$ in units of $P_0 = 2e|\Delta V| G_0$, with $\Delta V$ the applied voltage.} of the cavity correspond to the choices $a(x)=x$ and $a(x)=x(1-x)$ respectively in the Landauer-B\"uttiker theory \cite{land,fisherlee,butt}. 

The parameter $\alpha= N_1/N_2-1\geq 0$, kept fixed in the large $N_{1,2}$ limit, accounts for the asymmetry in the number of open electronic channels. For a symmetric cavity $\alpha=0$. Furthermore, it is well-known that in this setting the transport eigenvalues $\{\lambda_i\}$ are distributed according to a Jacobi ($\mathcal{J}$) ensemble \cite{jaccond1,jaccond2} with $V_\mathcal{J}(x)=(\alpha/2)\ln x$, implying an average density $\rho(\lambda)$ supported on $[\lambda_-,\lambda_+]=[\alpha^2/(\alpha+2)^2,1]$ (compare with Table \ref{table1}).

It was precisely in this quantum transport setting that Beenakker's formula (\textsf{B}) was first derived \cite{beenakkerPRL,beenakkerPRB}. It reads
\be
\mathrm{Var}(A)=\frac{1}{\beta \pi^2} \int_0^\infty\de k |F(k)|^2 k\ \tanh(\pi k)\ ,\label{formulaB}
\ee
where $F(k)=\int_{-\infty}^\infty\de x\ \mathrm{e}^{\mathrm{i}kx}\ a\left(\frac{1}{1+\mathrm{e}^x}\right)$. It is immediate to verify that \eqref{formulaB} is recovered from our \eqref{cov_fourier} upon setting $a(x)=b(x)$ and (\emph{crucially}) $\alpha=0$, implying $[\lambda_-,\lambda_+]=[0,1]$. If $\alpha\neq 0$ (asymmetric cavities), \eqref{formulaB} is not applicable and the variance of conductance and shot noise \emph{do} depend explicitly on $\alpha$ \footnote{From \eqref{cov_fourier} one gets $\mathrm{Var} G=(2/\beta)(\alpha+1)^2/(\alpha+2)^4$ and $\mathrm{Var}P=(2/\beta) \left(\alpha^4+2\alpha^2+4\alpha+2\right) (\alpha+1)^2/(\alpha+2)^8$.}, in agreement with \cite{savin3,savin2}. In addition, from \eqref{cov_fourier} one gets the covariance of conductance and shot-noise to leading order in the channel numbers
\be
\mathrm{Cov}(G,P)=-\frac{2}{\beta}\frac{\alpha^2(\alpha+1)^2}{(\alpha+2)^6}\ .  \label{covGP}
\ee
We have checked that this result is in agreement with the asymptotics of an exact finite-$N$ expression in \cite{savin2} valid for all $\beta$ (see also \cite{mezzadri} for $\beta=1,2,4$, \cite{kanz2} for $\beta=2$ and $\alpha=0$, \cite{savin} for $\beta=1,2$, and \cite{novaesrecurrence} for a different large-$N$ method).
The simple form \eqref{covGP} shows that for $N_{1,2}\gg1$ conductance and shot-noise are \emph{anticorrelated} for
any value of $\alpha$ to leading order in $N_{1,2}$. Moreover, for a symmetric ($\alpha = 0$) cavity the two observables are uncorrelated (for $\beta=2$ this was noticed in \cite{savin2}). Given that their joint (typical) distribution is Gaussian \cite{cunden}, they are also independent to leading order in $N$ for $\alpha=0$. As shown in Fig.~\ref{fig:cov}, at $\alpha^{\star}=(1+\sqrt{3})=2.73205\dots$ (independent of $\beta$) the anticorrelation between $G$ and $P$ is maximal and equal to $\mathrm{Cov}(G,P)\big|_{\alpha=\alpha^{\star}}=-1/54\beta$. Since simultaneous measurement of conductance and shot noise are possible \cite{exper}, a verification of this ``$1+\sqrt{3}$" effect might be within reach of current experimental capabilities.

\begin{figure}[t]
\centering
\includegraphics[width=1\columnwidth]{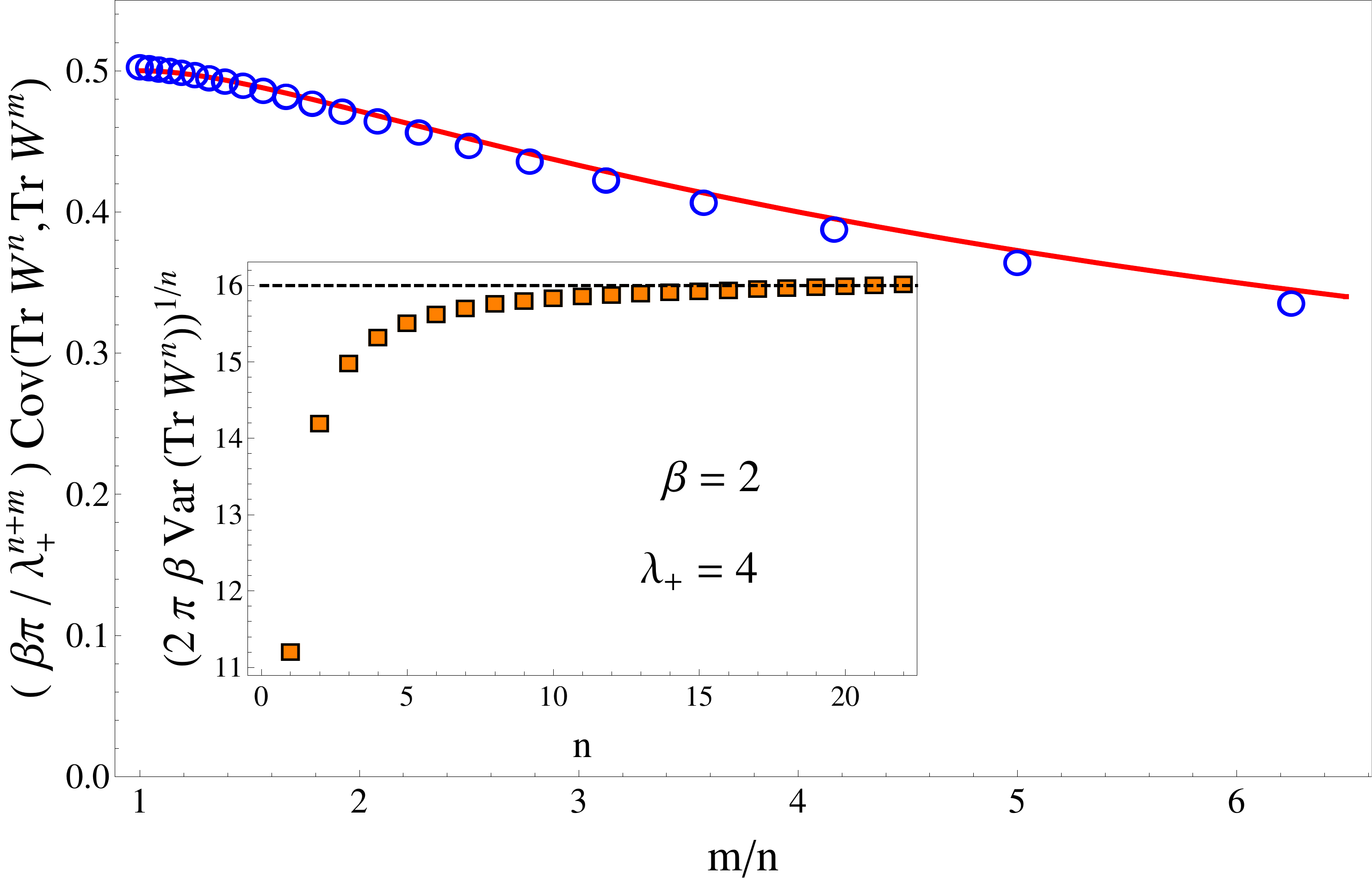}
\caption{(color online) The covariance $\mathrm{Cov}(\mathrm{Tr}\mathcal{W}^n,\mathrm{Tr}\mathcal{W}^m)$ as a function of $m/n$, where $\mathcal{W}$ is a complex Wishart matrix $(\beta=2)$. For the simulation we sampled $10^4$ $\mathcal{W}$ matrices of size $N=800$. Main: In the simulation (circles) $n$ is fixed to $50$ and $m$ varies. The solid curve is \eqref{asymp_covar}. Inset: the numerical simulations (squares) show the convergence of the rescaled variance to the limit value $\lambda_{+}^2=16$.} 
\label{fig:varTr}
\end{figure}

As a second example, we address the following question: what is the behavior of $\mathrm{Cov}\left(\mathrm{Tr}{\,\mathcal{H}^n},\mathrm{Tr}{\,\mathcal{H}^m}\right)$ as a function of $n$ and $m$ for an invariant matrix $\mathcal{H}$? Consider for simplicity an ensemble whose $\rho(\lambda)$ has support on the interval $\sigma=[0,\lambda_+]$, such as the Jacobi $\mathcal{J}$ or the Wishart-Laguerre $\mathcal{W}$ with $M=N$. In this case the conformal map reads $T(x)=\lambda_+/(x+1)$ and after a standard computation $\tilde{a}(k)=\lambda_{+}^n\mathrm{B}\left(\mathrm{i}k,n-\mathrm{i}k\right)$, where $B(a,b)=\int_0^1\de t\ t^{a-1}(1-t)^{b-1}$ is Euler's Beta function. For large $n$ and $m$, we may use the asymptotics $\mathrm{B}\left(\mathrm{i}k,n-\mathrm{i}k\right)=n^{-\mathrm{i}k}\left(\Gamma(\mathrm{i}k)+\mathcal{O}(n^{-1})\right)$ to get from \eqref{cov_fourier}
\be
\mathrm{Cov}\left(\mathrm{Tr}{\,\mathcal{H}^n},\mathrm{Tr}{\,\mathcal{H}^m}\right)\sim \frac{\lambda_+^{n+m}}{\beta \pi}\frac{\sqrt{nm}}{n+m}\ ,\label{asymp_covar}
\ee 
in perfect agreement with numerical simulations on $\mathcal{W}$ matrices (see Fig. \eqref{fig:varTr}). Setting $n=m$, we deduce the remarkable universal formula 
$
\lim_{n\to\infty} \left[ 2\pi\beta\,\mathrm{Var}(\mathrm{Tr}\,\mathcal{H}^n)\right]^{1/n} = \lambda_{+}^2\
$.
For the $\mathcal{W}$ ensemble this limiting value is $\lambda_{+}^2=16$ (see inset in Fig. \ref{fig:varTr}), while in the $\mathcal{J}$ case we get $\lambda_+=1$ recovering a result obtained in the context of the quantum transport problem (see \cite{vivoconductance}, Eq. (149), and \cite{novaes}). We now sketch the key steps of derivation of the general formula \eqref{cov_fourier} for $\sigma\neq\mathbb{R}$, treading in the same footsteps as \cite{beenakkerPRB}; mathematical details will be published elsewhere \cite{supp}. 

\textit{Derivation - } The starting point is Eq. \eqref{defcov11} together with \eqref{jpdf}. The crucial observation is that a change of variable $\lambda_i = T(x_i)$ induced by the conformal map $T(x)=(a x+b)/(c x+d)$ with $ad-cb\neq0$, transforms the original system into a new Coulomb gas of type \eqref{jpdf} at the \emph{same temperature} $\beta^{-1}$, with a modified potential $\tilde{V}(x)$ \footnote{The special case $T(x)=1/(1+x)$ of this general transformation was used in \cite{beenakkerPRL,beenakkerPRB} to derive \eqref{formulaB}.}. In these new variables, \eqref{defcov11} becomes
\be
\mathrm{Cov}(A,B)=\int_{\tilde{\Lambda}^N}\hspace{-2mm}\de\bm{x}\frac{1}{\tilde{\mathcal{Z}}}\mathrm{e}^{-\beta \tilde{E}(\bm{x})}A(T(\bm{x}))B(T(\bm{x}))-\avg{A}\avg{B}\ ,\label{defcov12}
\ee
with $ \tilde{E}(\bm{x})=-\sum_{i<j}\ln |x_i-x_j|+N\sum_i \tilde{V}(x_i)+\mathcal{O}(N)$.
Introducing the spectral density of the new system $\tilde{\rho}_N(x)=N^{-1}\sum_i{\delta(x-x_i)}$, \eqref{defcov12} can be reduced to the double integral
\be
\mathrm{Cov}(A,B)=-N^2\iint_{\tilde{\sigma}}\de x\de x^\prime \tilde{\mathcal{K}}_N(x,x^\prime)a(T(x))b(T(x^\prime))\ ,\label{generalcov1}
\ee
where $\tilde{\mathcal{K}}_N(x,x^\prime)=-\avg{\tilde{\rho}_N(x)\tilde{\rho}_N(x^\prime)}+\avg{\tilde{\rho}_N(x)}\avg{\tilde{\rho}_N(x^\prime)}$ is the two-point (connected) correlation function \footnote{Here $\avg{\cdot}$ stands for averaging with respect to $\tilde{P}_\beta(\bm{x})=\frac{1}{\tilde{\mathcal{Z}}}\exp(-\beta\tilde{E}(\bm{x}))$.}. We now denote $\tilde{\rho}(x)=\lim_{N\to\infty}\avg{\tilde{\rho}_N(x)}$ and $\tilde{\mathcal{K}}(x,x^\prime)=\lim_{N\to\infty} N^2\tilde{\mathcal{K}}_N(x,x^\prime)$. For a suitable choice of parameters $a,b,c,d$, the corresponding density $\tilde{\rho}$ is supported on $\tilde{\sigma}=(0,+\infty)$. In summary, the maps \eqref{T} are precisely constructed to achieve these goals - $\mathrm{i}.)$ the $2D$-Coulomb interaction (logarithmic) is preserved, and $\mathrm{ii}.)$ the support $\sigma$ is mapped into $\tilde{\sigma}=(0,+\infty)$ (this is possible whenever $\sigma$ has at most one point at infinity). If $\tilde{\rho}$ is supported on the positive half-line,
then the kernel reads \cite{beenakkerPRL,beenakkerPRB}
\be
\tilde{\mathcal{K}}(x,x^\prime)=\frac{1}{\pi^2}\frac{\de}{\de x}\frac{\de}{\de x^\prime}\ln\Big|\frac{\sqrt{x}-\sqrt{x^\prime}}{\sqrt{x}+\sqrt{x^\prime}}\Big|\ ,\label{kernelbeenakker}
\ee
valid for $x,x^\prime>0$. It is derived using the following two ingredients $\mathrm{i}.)$ the electrostatic integral equation for the density $\int_0^\infty\de x^\prime\tilde{\rho}(x^\prime)\ln |x-x^\prime|=\tilde{V}(x)$, which follows from a minimization argument of the energy of the Coulomb gas  \eqref{jpdf} \cite{Mehta}, and $\mathrm{ii.})$ the functional relation $\tilde{\mathcal{K}}(x,x^\prime)=\frac{1}{\beta}\frac{\delta\tilde{\rho}(\lambda)}{\delta \tilde{V}(\lambda^\prime)}$ \cite{beenakkerPRL,beenakkerPRB}, which descends from the definition $\avg{\tilde{\rho}_N(x)}=\int_{\tilde{\Lambda}^N}\de\bm{x}\frac{1}{\tilde{\mathcal{Z}}}\mathrm{e}^{-\beta \tilde{E}(\bm{x})}\tilde{\rho}_N(x)$ and the limit $N\to\infty$. Note that the universal $1/\beta$ behavior of \eqref{cov_fourier} is ultimately tracked back to this functional relation. As first noticed in \cite{beenakkerPRL}, the change of variables $x=\mathrm{e}^y$ and $x^\prime=\mathrm{e}^{y^\prime}$ makes the kernel \eqref{kernelbeenakker} translationally invariant and using standard results in Fourier space, the formula \eqref{cov_fourier} is readily established.
The main usefulness of the conformal map method (for $\sigma\neq\R$) is evident: the asymptotic kernel $\tilde{\mathcal{K}}(x,x^\prime)$ of the new gas \eqref{kernelbeenakker} becomes \emph{universal} (independent of details of the potential $\tilde{V}(x)$ and even of the edge points $\lambda_\pm$ of the original density $\rho(\lambda)$), yielding the fixed kernel $\varphi(k)$ in \eqref{cov_fourier}. Every surviving trace of the original ensemble is condensed in $\lambda_\pm$, which have now been moved inside the argument of the linear statistics.

\textit{Conclusions -}  In summary, we derived a universal formula \eqref{cov_fourier} for the covariance $\mathrm{Cov}(A,B)$ of two smooth linear statistics for one-cut random matrix models. Remarkably, the only dependence on the specific choice of the confining potential $V(\lambda)$ is through the edge points $\lambda_\pm$ of the average spectral density $\rho(\lambda)$. This is a consequence of universality of the (smoothed) two-point kernel, a classical result in Random Matrix Theory. 
It is worth to stress that a prominent role is played by the support $\sigma$ of the limiting density $\rho(\lambda)$ and \emph{not} by the support $\Lambda^N$ of the jpd $\mathcal{P}_{\beta}(\bm{\lambda})$. A welcome feature of \eqref{cov_fourier} is that the covariance of $A(\bm\lambda)$ and $B(\bm\lambda)$ may vanish, a possibility that obviously cannot materialize for the \emph{variance} of a nontrivial linear statistics. The consequence is that some linear statistics of the same ensemble may be \emph{uncorrelated} to leading order in $N$, despite being functions of \emph{strongly correlated} eigenvalues (see \cite{vivoluque, vivoluque2,novaesdecorr} for other occurrences of this phenomenon). A joint Gaussian behavior - as already detected in a few cases \cite{cunden, cunden2} - would then also imply \emph{independence}.

We gave two examples of application. The first one is borrowed from the theory of electronic transport in chaotic cavities, yielding a neat formula \eqref{covGP} for the covariance of conductance and shot noise, and its previously unnoticed non-monotonic behavior (see Fig. \ref{fig:cov}). As a second example, we provide the asymptotic formula \eqref{asymp_covar} for the correlation of traces of higher matrix powers. We leave further applications for a forthcoming work \cite{supp}. In the future, it will be interesting to search for extension of formula \eqref{cov_fourier} to multi-cuts matrix models \cite{freikanz,akemann,bonnet} as well as to the case of non-hermitian random matrices \cite{forrestercomplex}. A thorough investigation of non-smooth linear statistics (see e.g. \cite{pastur}) is also very much called for in the context of number variance and index problems \cite{pastur,scard,fogler,variance1}. It should also be possible to study the covariance of linear statistics for the biorthogonal case \cite{duits} and, in general, establishing a Central Limit Theorem for \emph{joint} fluctuations of linear statistics is a task whose accomplishment may turn out to be relevant for several applications. 

\textit{Acknowledgments - } We are indebted to G. Akemann, C. W. J. Beenakker, P. Facchi, M. Novaes and D. Savin for valuable correspondence and much helpful advice. PV acknowledges support from Labex-PALM (Project Randmat).

\end{document}